\documentclass{article}
\usepackage{amsmath,amssymb,graphicx,amsfonts}
\usepackage{algorithmic}
\usepackage{algorithm}
\usepackage{tikz,xcolor}
\usetikzlibrary{backgrounds,decorations.pathmorphing,arrows}

\def\ZZ{\mathbb Z}

\def\cC{\mathcal C}

\def\cG{\mathcal G}

\def\cN{\mathcal N}

\def\cE{\mathcal E}

\def\b1{\mathbf 1}

\newcommand{\nop}[1]{}

\newcommand{\qed}{\hfill$\square$\bigskip}
\newcommand{\raf}[1]{(\ref{#1})}
\newcommand{\proof}{\noindent {\bf Proof}.~~}

\newcommand{\hide}[1]{}

\newtheorem{theorem}{Theorem}

\newtheorem{lemma}{Lemma}
\newtheorem{corollary}{Corollary}
\newtheorem{proposition}{Proposition}

\title{A Polynomial Delay Algorithm for Generating Connected Induced Subgraphs of a Given Cardinality}

\author{
Khaled Elbassioni\thanks{Department of Electrical Engineering and Computer Science, Masdar Institute of Science and Technology, P.O.Box 54224, Abu Dhabi, UAE;
(kelbassioni@masdar.ac.ae)}
}

\begin{document}
\date{}
\maketitle
\begin{abstract}
We give a polynomial delay algorithm, that for any graph $G$ and positive integer $k$,  enumerates all connected induced subgraphs of $G$ of order $k$. Our algorithm enumerates each subgraph in at most $O((k\min\{(n-k),k\Delta\})^2(\Delta+\log k))$ and uses linear space $O(n+m)$, where $n$ and $m$ are respectively the number of vertices and edges of $G$ and $\Delta$ is the maximum degree.  
\end{abstract}

\section{Introduction}\label{s1}
Let $G=(V,E)$ be an undirected graph of $|V|=n$ vertices and $m=|E|$ edges.
Given an integer $k\in\ZZ_+$, we consider the problem GEN$(G;k)$ of enumerating (or generating) all subsets $X\subset V$ of vertices such that $|X|=k$ and the subgraph $G[X]$ induced on $X$ is connected. Let us denote the family of such vertex sets by $\cC(G;k)$.  

Typically, the size of $\cC(G;k)$ is exponentially large in $k$. In fact, it was shown in \cite{U} that there exists a family of connected graphs $G$ with maximum degree $\Delta<\frac{2n}{k}$, for which $|\cC(G;k)|\ge n(\frac{\Delta}{2})^{k-1}$. On the other hand, an upper bound\footnote{{\color{red} Note that the factor of $n$ in this bound was mistakenly dropped in \cite{U}; this was propagated in an earlier version of the current paper \cite{E15} resulting in the omission of a factor of $\log n$ in Corollary~\ref{c-main}.}} of $n\cdot\frac{(e\Delta)^k}{(\Delta-1)k}$ on the size of $\cC(G;k)$ was also given in \cite{U}.

An enumeration algorithm for $\cC(G;k)$ is said to be {\it output (or total) polynomial} if the algorithm outputs all the elements of $\cC(G;k)$ in time polynomial in $n$ and $|\cC(G;k)|$. Avis and Fukuda \cite{AF96} introduced the {\it reverse search method} for enumeration, and used it to solve, among several other problems, the problem of enumerating all connected induced subgraphs of size {\it at most} $k$.
Uehara \cite{U} noted that such an algorithm is total polynomial for enumerating $\cC(G;k)$ when $k$ is a fixed constant. In fact, since the algorithm of \cite{AF96} enumerates the families $\cC(G;i)$ for all $1\le i \le k$, the total the running time of the algorithm is bounded by (see \cite{U}):
\begin{equation}\label{rt}
O\left(n+m+\sum_{i=1}^{k-1}|\cC(G;i)|+k^2|\cC(G;k)|\right),
\end{equation}
which is upper bounded by $O\left(m+n\cdot\frac{(e\Delta)^k}{(\Delta-1)}k^2\right)$, using the upper bound on $|\cC(G;k)|$ mentioned above.

However, we note that such a bound is not polynomial, when $k$ is part of the input\footnote{An analogy can be drawn from the enumeration of face lattices of polytopes, where enumerating all faces of dimension at most $k$ is solvable in total polynomial time, while enumerating faces of dimension exactly $k$ includes the well-known {\it vertex enumeration problem}, whose complexity is an outstanding open question. This discrepancy is due to the fact that, there are polytopes (the so-called {\it fat-lattice} polytopes, see e.g. \cite{ABS97, D83}) where the total number of faces is exponentially large in the number of facets and vertices}.
In fact, considering the lower bound example mentioned above, we observe that,
when $k=n-\Delta$, the size of $\cC(G;k)$ is at most $\binom{n}{k} < n^\Delta$, while 
for all $i<\frac{2n}{\Delta}$, $|\cC(G;i)|\ge n(\frac{\Delta}{2})^{i-1}$. Thus for $i=\frac{2n}{\Delta}-1$, the total running time in \raf{rt} will be at least $n(\frac{\Delta}{2})^{\frac{2n}{\Delta}-2}$. Setting, for instance, $\Delta=n^{\epsilon}$ for some $\epsilon\in(0,\frac{1}{2})$, and assuming $n$ is large enough, we get that $i=2n^{1-\epsilon}-1<k=n-n^\epsilon$, and thus the running time in \raf{rt} is at least 
$$n^{(2n^{1-\epsilon}-2)(\epsilon-\frac{1}{\log n})+1}>|\cC(G;k)|^{(2n^{1-2\epsilon}-2n^{-\epsilon})(\epsilon-\frac{1}{\log n})}=|\cC(G;k)|^{n^{\Omega(1)}}.$$    
Thus, the algorithm suggested in \cite{U} is {\it not} a total polynomial algorithm.  

\medskip

An enumeration algorithm for GEN$(G;k)$ is said to have a {\it polynomial delay} of $p(n,m,k)$ if it outputs all the elements of $\cC(G;k)$, such that the running time between any two successive outputs is at most $p(n,m,k)$.  The algorithm is said to use {\it linear space}, if the total space required by the algorithm (excluding the space for writing the output)  is $O(n+m)$. Recently, Karakashian et al. \cite{KCH13} gave an algorithm with delay $O(\Delta^k)$ for solving GEN$(G;k)$. 

In this short note, we give a bound that is {\it polynomial} in $k$.

\begin{theorem}\label{main}
	There is an algorithm for solving GEN$(G;k)$, for any graph $G=(V,E)$ and integer $k\in\ZZ_+$, with polynomial delay $O((k\min\{(n-k),k\Delta\})^2(\Delta+\log k))$ and space $O(n+m)$.
\end{theorem}

We remark that the polynomial delay bound can be improved if we do not insist on using polynomial space (that is, the space used by the algorithm may depend polynomially on $|\cC(G;k)|$); see Corollary~\ref{c-main} below.

Our proof of Theorem~\ref{main} is also based on using the the reverse search method \cite{AF96}. In fact we will consider more generally the {\it supergraph method} for enumeration, which can be thought of as a generalization of the reverse search method.
This method will be briefly explained in the next section. Then we prove Theorem~\ref{main} in Sections~\ref{Neighborhood} and \ref{StrongConnectivity}. 

\medskip

The problem of enumerating of connected induced subgraphs of size $k$ arises in several applications, such as keyword search over RDF graphs in Information Retrieval, and consistency analysis in Constrained Processing; see, e.g., \cite{EB11,KCH13} and the references therein.

\section{The Supergraph Approach}\label{supergraph}
This technique generally works by building and traversing a directed (super)graph
$\cG=(\cC(G;k),\cE)$, defined on the family $\cC(G;k)$.
The arcs of $\cG$ are defined by a {\it neighborhood function}
$\cN:\cC(G;k) \to 2^{\cC(G;k)}$, that to any
$X\in\cC(G;k)$ assigns a set of its successors $\cN(X)$ in $\cG$.
A special node $X_0\in\cC(G;k)$ is identified from which all
other nodes of $\cG$ are reachable. The algorithm works by
traversing, either, in {\it depth-first} or {\it breadth-first} search order, the nodes of
$\cG$, starting from $X_0$. If $\cG$ is {\it strongly connected} then
$X_0$ can be any node in $\cC(G;k)$.

To avoid confusion, in the following, we will distinguish the vertices of $G$ and $\cG$ by referring to them as {\it vertices} and {\it nodes}, respectively.

\medskip

The following fact is known about this approach (see e.g.
\cite{AF96,BEGM-CRM,JYP88,SS02}):

\begin{proposition} \label{p1}
	Consider the supergraph $\cG$ and suppose that 
	\begin{itemize}
		\item[(i)] $\cG$ is strongly connected;
		\item[(ii)] a node $X_0$ in $\cG$ can be found in time $t_0(n,m,k)$;
		\item[(iii)] for any node $X$ in $\cG$, $|\cN(X)|\le N(n,m,k)$ and we can generate $\cN(X)$ with delay $t(n,m,k)$;
	\end{itemize}
	then $\cC(G;k)$ can be generated with delay $O(\max\{t_0(n,m,k),(t(n,m,k)+\log |\cC(G;k)|)\cdot N(n,m,k)\}$ and space $O(n+m+k|\cC(G;k)|)$. 
	If instead of (i), 
	\begin{itemize}
		\item[(iv)] there is a function $f:\cC(G;k)\setminus \{X_0\}\to\cC(G;k)$, that for every node $X\neq X_0$ in $\cG$, identifies (in a unique way), in time $t_1(n,m,k)$, a node $X'=f(X)$ in $\cG$ such that $X\in\cN(X')$, and $f$ satisfies the following {\it acyclicity property}: there exist no $X_{\ell+1}=X_1,X_2,\ldots,X_\ell\in\cC(G;k)$ such that $X_i=f(X_{i+1})$ for $i=1,\ldots\ell$,
	\end{itemize}
	then $\cC(G;k)$ can be generated with delay $O(\max\{t_0(n,m,k),(t(n,m,k)+t_1(n,m,k))\cdot  N(n,m,k)\}$ and space $O(n+m)$.
\end{proposition}

The proof is straightforward. For the first claim, we essentially traverse a {\it breadth-first} search (BFS) tree on $\cG$, starting from node $X_0$. We maintain a {\it balanced binary search tree} (BST) on the elements generated, sorting them, say, according to some {\it lexicographic order}. We also keep a {\it queue} of all elements that have been generated but whose neighborhoods have not been yet explored. When processing a node $X$ in the queue, we output $X$ and then generate all its neighbors in $\cG$, but only insert in the queue a neighbor $X'$ if it has not been yet stored in the BST. 

To achieve the second claim, we instead traverse a {\it depth-first} search (DFS) tree on $\cG$, starting from node $X_0$. (This is essentially the reverse search method) When traversing a node $X$, we generate all neighbors of $X$ in $\cG$, but only proceed the search on a neighbor $X'$, if $X=f(X')$ is the unique "parent" defined in (iv). The fact that $f$ is a function satisfying the acyclicity property implies that all the nodes in $\cG$ are processed exactly once (since for any node $X$ in $\cG$ there is a unique path leading to $X_0$). In order to obtain  the claimed delay bound,
we output a node $X$ just after the first visit to it, if the depth of
$X$ in the tree is odd (assume the root $X_0$ has depth 1), or $X$ is a leaf; otherwise, $X$ is output just before coming back to the parent. (Note that  this way of distributing the outputs over time ensures that the delay between two successive outputs is not more than the time to generate the neighborhoods of at most two nodes. Indeed, if $X$ has odd depth or $X$ is a leaf, then the last node output before $X$ would be the closet ancestor of $X$ at odd depth; if $X$ has even depth, then the last node output before $X$ would be the closest node with even depth (or a child leaf if no such node exists) on the right-most path descending from $X$.)        
Note that we do not need to store the history along any search path since the parent of any node can be generated efficiently. 

\medskip

Clearly, we may and will assume in the rest of the paper that the graph $G$ is connected.
In the next section, we will prove that this assumption implies that the supergraph $\cG$ in our case is strongly connected (and in fact has diameter $\delta\le n-k$). Let us now set the other parameters in Proposition~\ref{p1}, corresponding to our problem GEN$(G;k)$. We assume an order on the vertices of $G$, defined by an arbitrary DFS tree, which also naturally defines a lexicographic order "$\preceq$" on the vertex sets. Clearly, the lexicographically smallest node $X_0\in\cC(G;k)$ can be found in time $t_0(n,m,k)=O(k\Delta)$ by traversing the DFS tree starting from the smallest vertex in $G$ and processing vertices in DFS order until exactly $k$ vertices are visited.

Given a node $X\in\cC(G;k)$, we have, with the neighborhood definition given below, $|\cN(X)|\le k\cdot \min\{(n-k),k\Delta\}$, and each element in $\cN(X)$ can be generated in time\footnote{This can be achieved by using a simple disjoint-set data structure; see e.g.  Theorem~21.1 in \cite{CLRS09}. We remark that this bound can be improved through the use of more sophisticated data structures.} $t(n,m,k)=O(k (\Delta+\log k))$. These bounds, together with the upper bound $|\cC(G;k)|\le n\cdot \frac{(e\Delta)^k}{(\Delta-1)k}$ imply that the set $\cC(G;k)$ can be generated with polynomial delay $O(k\cdot\min\{(n-k),k\Delta\}\cdot(k(\Delta+\log k)+\log n))$ and space $O(n+m+k|\cC(G;k)|)$. 

In order to achieve a polynomial space bound (independent of $|\cC(G;k)|$), we will need a stronger claim, namely that every node $X\in\cC(G;k)\setminus\{X_0\}$ is reachable from $X_0$ by a {\it monotonically increasing} lexicographically ordered sequence of nodes. This claim will be proved for any DFS ordering on the vertices of $G$, and allows us to identify the parent $X'$ of any node $X$ by defining, for instance, $X'=X\cup\{u\}\setminus\{v\}$, where $u\in V\setminus X$ and $v\in X$ are respectively the smallest and largest vertices (in the DFS order) such that $X'\prec X$ and the graph $G[X\cup\{u\}\setminus\{v\}]$ is connected. Note that finding $X'$ can be done in time $t_1(n,m,k)\le t(n,m,k)\cdot \min\{(n-k),k\Delta\}$.

\section{The Neighborhood Operator for $\cC(G;k)$}\label{Neighborhood}
For a set $X\in\cC(G;k)$, it is natural
to define the neighbors of $X$ as those which are obtained from $X$ by exchanging one vertex:
$$\cN(X)=\{X'\in\cC(G;k):~X\cap X'=k-1\}.$$
It is worth comparing our neighborhood definition to the one suggested in \cite{AF96} for generating all connected induced subgraphs of size {\it at most} $k$. In the latter definition, two sets $X,X'\subseteq V$ are neighbors if they {\it differ} in exactly one vertex. The claim of strong connectivity follows immediately from the simply facts that if $X\subset V$ is such that $G[X]$ is connected then there is a vertex $u\in X$ such that $G[X\setminus\{u\}]$ is also connected, and similarly, there is a vertex $u\in V\setminus X$ such that $G[X\cup\{u\}]$ is connected. 

\section{Strong Connectivity} \label{StrongConnectivity}

We prove first that the supergraph $\cG$ is strongly connected. 
\begin{lemma}\label{l-main}
	Let $X,Y$ be two distinct elements of $\cC(G;k)$. Then there exist vertex sets $X_1,X_2,\ldots,X_\ell\in\cC(G;k)$ such that $X_1=X$, $X_\ell=Y$, $\ell\le n-k+1$, and for $i=1,\ldots,\ell-1$, $X_{i+1}\in\cN(X_{i})$.  
\end{lemma}
\proof
Let $d(Z,Z')$ be the (shortest) distance between the two vertex sets $Z,Z'$ in $G$. Suppose we have already constructed $X_{i}$. We consider two cases.

\smallskip

\noindent{\it Case 1.} $d(X_i,Y)>0$ (and hence $X_i\cap Y=\emptyset$). Let $u_0,u_1,\ldots, u_r$ be the ordered sequence of vertices on the shortest path between $X_i$ and $Y$ in $G$, where $u_0\in X_i$ and $u_r\in Y$. Let $T$ be a spanning tree in $G[X_i]$. Then $T$ has a leaf $v\ne u_0$. We define $X_{i+1}=X_i\cup\{u_1\}\setminus\{v\}$. By construction, $X_{i+1}\in\cC(G;k)$, and $d(X_{i+1},Y)<d(X_i,Y)$. Since $d(X_1,Y)\le n-2k+1$, we will arrive at case 2 after at most $n-2k+1$ iterations.

\smallskip

\noindent{\it Case 2.} $d(X_i,Y)=0$. Then there exists a vertex $z\in X_i\cap Y$. Let $C(X_i;z)$ be (the vertex set of) the connected component containing $z$ in $G[X_i\cap Y]$. We claim that there exists a vertex set $X_{i+1}\in\cN(X_i)$ such that $|C(X_{i+1};z)|>|C(X_i;z)|$. Indeed, let us contract $C(X_i;z)$ into a single vertex $w$ and denote the new graph by $G'$. Then, by the connectivity of $G[Y]$, there is an edge $\{w,u\}$ in the graph $G'[Y\cup\{w\}\setminus C(X_i;z)]$, where $u\in Y$ ($u\ne w$). Similarly, the graph $G'[X_i\cup\{w\}\setminus C(X_i;z)]$ is connected and hence has a spanning tree $T$. Let $v\ne w$ be a leaf in $T$. Then $X_{i+1}=X_i\cup\{u\}\setminus \{v\}$ satisfies the claim. This claim implies that in at most $k-1$ iterations of case 2, we will have $X_{i+1}=Y$. 
\qed

In view of Proposition~\ref{p1}, Lemma~\ref{l-main} implies that all the elements of $\cC(G;k)$ can be enumerated with polynomial delay. 
\begin{corollary}\label{c-main}
	There is an algorithm for solving GEN$(G;k)$, for any graph $G=(V,E)$ and integer $k\in\ZZ_+$, with polynomial delay $O(k\cdot\min\{(n-k),k\Delta\}\cdot(k(\Delta+\log k)+\log n))$. \end{corollary}

In order to prove that GEN$(G;k)$ can be solved also with polynomial space, we need the following result.

\begin{lemma}\label{l-main2}
	Consider the lexicographic ordering "$\preceq$" on $\cC(G;k)$ defined by a DFS order on the vertices of $G$. Let $X$ be any element in $\cC(G;k)$ that is not lexicographically smallest. Then there is an $X'\in\cC(G;k)$ such that $X'\in\cN(X)$ and $X'\prec X$.  
\end{lemma}
\proof
Let us denote the lexicographically smallest element of $\cC(G;k)$ by $X_0$. Since $X_0\prec X$, it holds that $w:=\min_{u\in X_0\setminus X}u<z:=\min_{u\in X\setminus X_0}u$. We use the following simple property of the  DFS tree: for $x<y$, the DFS-tree walk  (that is, the sequence of vertices visited on the way in the DFS order) from $x$ to $y$ contain only nodes $z\le y$.  
We consider a number of cases:

\smallskip

\noindent{\it Case 1.} $z$ is not a {\it cut-vertex} in $G[X]$ (that is, $G[X]-z$ is connected). We consider two subcases.

\smallskip

\noindent{\it Subcase 1.1.} The DFS-tree walk from $w$ to $z$ enters $X$ at a vertex $x\neq z$ through an edge $\{y,x\}$. Thus $x\in X$, $y\not\in X$ and $y<z$. Since $z$ is not a cut vertex, there is a spanning tree in $G[X]$ which has $z$ as a leaf. Then $X'=X\cup\{y\}\setminus\{z\}$ satisfies the claim. 

\smallskip

\noindent{\it Subcase 1.2.} The DFS-tree walk from $w$ to $z$ enters $X$ at $z$ through an edge $\{y,z\}$. If all vertices $u\in X\setminus \{z\}$ satisfy $u>z$, then let $x\neq z$ be a leaf in a spanning tree in $G[X]$, and set $X'=X\cup\{y\}\setminus\{x\}$ which satisfies the claim.   
Otherwise, there is a vertex $x$ in $X\cap X_0$ that precedes $z$ in the DFS order. Then $x$ necessarily precedes $w$ (otherwise we are in case 1.1, as the DFS walk from $w$ to $x$ would enter $X$ through an edge $\{u,v\}$, where $u<x<z$), and the DFS walk between $x$ and $w$ must exit $X$ through some edge $\{u,v\}$ with $u\in X$ and $v\not\in X$ such that $v\le w<z$. Then setting $X'=X\cup\{v\}\setminus\{z\}$ satisfies the claim.  

\smallskip

\noindent{\it Case 2.} $z$ is a {\it cut-vertex} in $G[X]$. Let $X_1$ be the vertex set of the connected component in $G[X]-z$ containing the smallest vertex $u$ in $X\setminus\{z\}$, and let $X_2$ be the vertex set of any other connected component in $G[X]-z$. Let $v\ne z$ be a leaf in a spanning tree in $G[X_2\cup\{z\}]$. We consider two subcases.

\smallskip

\noindent{\it Subcase 2.1.} The DFS-tree walk from $u$ to $v$ does not go through $z$. Then it must be the case that the walk leaves $X_1$, and hence $X$, through an edge $\{x,y\}$, where $x\in X_1$ and $y\not\in X$. Since $y<v$, the set $X'=X\cup\{y\}\setminus\{v\}$ satisfies the claim.

\smallskip

\noindent{\it Subcase 2.2.} The DFS-tree walk from $u$ to $v$ does go trough $z$. Then $z<v$ and the DFS walk from $w$ to $z$ enters $X$ at a vertex $x\ne v$ through and edge  $\{y,x\}$, where $y\not\in X$. Since $y< z<v$, we can set $X'=X\cup\{y\}\setminus\{v\}$ to satisfy the claim.  
\qed

We note that the conclusion of Lemma~\ref{l-main2} is not true if we replace the DFS order of the vertices by an arbitrary order. Consider for instance, the case when $G$ is a path with vertices numbered $1,2,\ldots,k,n,,k+1,\ldots,n-1$, in the order they appear on the path, where $n>2k+1$. Then the set $X=\{k+1,k+2,\ldots,2k\}$ has two neighbors, namely, $X'=\{n,k+1,k+2,\ldots,2k-1\}$ and $X''=\{k+2,k+2,\ldots,2k+1\}$, both of which are lexicographically larger than $X$. 

\bigskip

\section*{Acknowledgements}
I thank Shady Elbassuoni for bringing this problem to my attention, and the anonymous reviewer for the careful reading and many  valuable remarks.



\end{document}